\documentclass[twocolumn,showpacs,aps,prl,superscriptaddress]{revtex4}    
\usepackage{graphicx}
\usepackage{dcolumn}
\usepackage{amsmath}
\usepackage{epsfig}    
\usepackage{rotating}

\input pubboard/babarsym

\newcommand{\BABARPubYear}    {01}
\newcommand{\BABARPubNumber}  {19}
          
\newcommand{\SLACPubNumber} {9102}

\def\figurebox#1#2#3{%
    \def\arg{#3}%
    \ifx\arg\empty
    {\hfill\vbox{\hsize#2\hrule\hbox to #2{\vrule\hfill\vbox to #1{\hsize#2\vfill}\vrule}\hrule}\hfill}%
    \else
    {\hfill\epsfbox{#3}\hfill}%
    \fi}
 
\long\def\inst#1{\par\nobreak\kern 4pt\nobreak
    {\it #1}\par\vskip 10pt plus 3pt minus 3pt}

\begin{document}

\preprint{\babar-PUB-\BABARPubYear/\BABARPubNumber}
\preprint{SLAC-PUB-\SLACPubNumber}
 
\begin{flushleft}
\babar-PUB-\BABARPubYear/\BABARPubNumber\\
SLAC-PUB-\SLACPubNumber\\[10mm]
\end{flushleft}

\title{
{\large \bf \boldmath Search for the Rare Decays 
$B \rightarrow K \ell^+ \ell^-$ and  $ B \rightarrow K^{\ast} \ell^+ \ell^-$}
}

%
\author{B.~Aubert}
\author{D.~Boutigny}
\author{J.-M.~Gaillard}
\author{A.~Hicheur}
\author{Y.~Karyotakis}
\author{J.~P.~Lees}
\author{P.~Robbe}
\author{V.~Tisserand}
\affiliation{Laboratoire de Physique des Particules, F-74941 Annecy-le-Vieux, France }
\author{A.~Palano}
\author{A.~Pompili}
\affiliation{Universit\`a di Bari, Dipartimento di Fisica and INFN, I-70126 Bari, Italy }
% ...
\collaboration{The \babar\ Collaboration}
\noaffiliation
         
\date{January 4, 2002}

\begin{abstract}
We present results from a search for the flavor-changing neutral current
decays $B\to K\ell^+\ell^-$ and $B\to K^*\ell^+\ell^-$,
where $\ell^+\ell^-$ is either an $e^+e^-$ or $\mu^+\mu^-$ pair.
The data sample comprises $22.7\times 10^6$
$\Upsilon(4S)\to \BB$ decays collected with the \babar\ detector at the \pep2 \BF. 
We obtain 
the 90\% C.L.~upper limits  
${\mathcal B}(B\to K\ell^+\ell^-)< 0.50\times 10^{-6}$ and 
${\mathcal B}(B\to K^*\ell^+\ell^-)<2.9\times 10^{-6}$, close to
Standard Model predictions for these branching fractions. We have
also obtained limits on the lepton-family-violating 
decays $B\to Ke^{\pm}\mu^{\mp}$ and $B\to K^{*}e^{\pm}\mu^{\mp}$.

\end{abstract}
 
\maketitle
\par    
   


The flavor-changing neutral current decays 
$B\to K\ell^+\ell^-$ and $B\to K^*(892)\ell^+\ell^-$, 
where $\ell^{\pm}$ is 
a charged lepton, are highly suppressed in the Standard 
Model, with branching fractions 
predicted
to be of order 
$10^{-7}-10^{-6}$~\cite{bib:TheoryA,bib:TheoryB}. The dominant contributions arise at
the one-loop level and 
are known as electroweak penguins.  
Besides probing Standard Model loop effects, these rare decays are important
because their rates and kinematic distributions are sensitive
to new, heavy particles---such as those predicted by 
supersymmetric models---that can appear virtually in the 
loop~\cite{bib:TheoryA,bib:TheoryB}. 

The Standard Model predictions for $B\to K^{(*)}\ell^+\ell^-$
include three main contributions: the electromagnetic (EM) penguin,
the $Z$ penguin, and the $W^+W^-$ box diagram. 
Evidence for the EM penguin amplitude has been obtained
from the observation of $B\to K^*\gamma$
and inclusive $B\to X_s\gamma$, where $X_s$ is any hadronic system with 
strangeness~\cite{bib:CLEOKstargam,bib:CLEOXsgam}.

Calculations of decay rates for $B\to K^{(*)}\ell^+\ell^-$ 
based on the Standard Model 
have significant uncertainties
due to strong interactions. 
For example, 
Ali {\it et al.}~\cite{bib:TheoryA} predict
${\cal B}(B\to K\ell^+\ell^-)=(0.57^{+ 0.17}_{-0.10})\times 10^{-6}$ for both $e^+e^-$ and $\mu^+\mu^-$
final states, 
${\cal B}(B\to K^* e^+e^-)=(2.3^{+0.7}_{-0.5})\times 10^{-6}$, and 
${\cal B}(B\to K^* \mu^+\mu^-)=(1.9^{+0.5}_{-0.4})\times 10^{-6}$. 
The contribution of the EM penguin amplitude to
$B\to K^{*}\ell^+\ell^-$ is particularly strong 
at low values of $m_{\ell^+\ell^-}$, giving a larger rate for $B\to K^* e^+e^-$ than 
for $B\to K^* \mu^+\mu^-$.
 
We search for the following decays:
$B^+\to K^+\ell^+\ell^-$, 
$B^0\to K_S^0\ell^+\ell^-$,
$B^+\to K^{*+}\ell^+\ell^-$, and 
$B^0\to K^{*0}\ell^+\ell^-$, where
$K^{*0}\to K^+\pi^-$, $K^{*+}\to K_S^0\pi^+$,
$K_S^0\to\pi^+\pi^-$, and
$\ell$ is either an 
$e$ or $\mu$.
We also search
for the lepton-family-violating decays
$B\to K^{(*)}e^{\pm}\mu^{\mp}$.
Throughout this paper, charge-conjugate modes are implied.


The data used in the analysis were collected with the \babar\ detector
at the \pep2\ storage ring at the Stanford Linear Accelerator Center
during 1999-2000.
We analyzed a 20.7 \invfb\ data sample taken on the 
$\Upsilon(4S)$ resonance
consisting of 
$(22.7\pm0.4)\times 10^6$ $\Upsilon(4S)\to \BB$ events.

This search relies primarily on the charged-particle tracking and
particle-identification capabilities of the  
\babar\ detector~\cite{bib:babarNIM}.
Charged particle tracking is provided by a five-layer silicon
vertex tracker (SVT) and a 40-layer drift chamber (DCH). 
The DIRC, a Cherenkov ring-imaging particle-identification system,
is used for charged hadron identification.
Electrons are identified using
the electromagnetic calorimeter (EMC), which comprises 6580 thallium-doped CsI
crystals. These systems are mounted inside a 1.5 T solenoidal
superconducting magnet. Muons are identified in the instrumented flux 
return (IFR), in which resistive plate chambers are interleaved
with the iron plates of the magnet flux return.


We extract the signal using the kinematic variables
$m_{\rm ES}= \sqrt{E_{\rm b}^{*2} - (\sum_i {\bf p}^*_i)^2}$
and $\Delta E= \sum_i\sqrt{m_i^2 + {\bf p}_i^{*2}}- E_{\rm b}^*$,
where $E_{\rm b}^*$ is the beam energy in the $e^+e^-$ rest (c.m.) frame,
${\bf p}_i^*$ is the c.m.~momentum of daughter particle $i$ in the
$B$ meson candidate, and $m_i$ is the mass of particle $i$. 
For signal events, $m_{\rm ES}$ peaks at the $B$ meson mass with 
a resolution of about 2.5 MeV$/c^2$ and $\Delta E$ peaks near zero,  
indicating that the candidate system of particles has total energy consistent with
the beam energy in the c.m.~frame.
To prevent bias in the analysis, we optimized the
event-selection criteria using Monte Carlo samples:
we did not look at the data in the signal
region or in the sidebands that were used to measure 
the background until these criteria were fixed. 
Signal efficiencies were
determined using the Ali {\it et al.} model~\cite{bib:TheoryA}.

We select events that have at least four charged tracks,
the ratio $R_2$ of the second and zeroth Fox-Wolfram moments~\cite{bib:FoxWolfram}
less than 0.5, and 
two 
oppositely charged leptons with 
momentum
\mbox{$p > 0.5 \ (1.0) \ {\rm GeV}/c$} for $e$ ($\mu$) 
candidates.  
Electron-positron pairs consistent with photon conversions
in the detector material are vetoed.
We require charged kaon candidates to be identified as kaons and
the charged pion in $K^{*}\to K\pi$  not to be identified 
as a kaon.
For \mbox{$B\rightarrow K^{\ast} \ell^+\ell^-$}, 
we require the mass of the $K^\ast$ candidate to 
be within \mbox{75$~\mevcc$} of the mean $K^\ast(892)$ mass.  
$K^0_S$ candidates are reconstructed from two oppositely charged tracks 
that form a good vertex
displaced from the primary vertex by at least 1 mm.

\begin{figure}[!tb]
 \begin{center}
   \includegraphics[width=\linewidth]{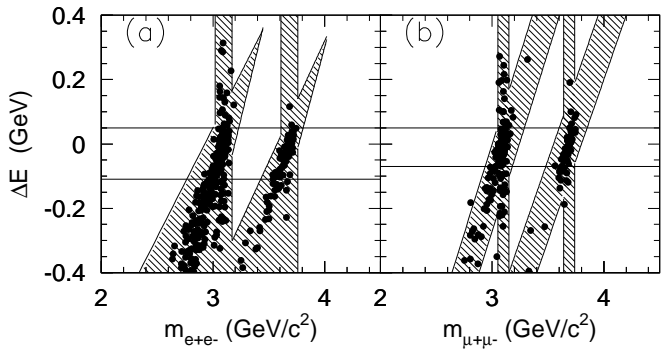}
  \end{center}
  \vspace{-0.5cm}
\caption[Definition of the charmonium veto region.]
{\label{fig:charmoniumveto}
Charmonium veto in the $\Delta E$ vs.~$m_{\ell^+\ell^-}$ plane
for (a) $B\to K^{(*)} e^+e^-$
and (b) $B\to K^{(*)} \mu^+\mu^-$. Hatched regions are vetoed.  
The dots correspond to a Monte Carlo simulation of 
$B\to J/\psi(\to\ell^+\ell^-)K$ and $B\to\psi(2S)(\to\ell^+\ell^-)K$.
Most signal events would lie in the $\Delta E$ region
between the horizontal lines.
}
\end{figure}

The decays \mbox{$B \to J/\psi(\to \ell^+\ell^-)K^{(*)}$} and
\mbox{$B \to \psi(2S)(\to \ell^+\ell^-)K^{(*)}$} have identical 
topologies to signal events.
These backgrounds
are suppressed by applying a veto in 
the $\Delta E$ vs.~$m_{\ell^+\ell^-}$ plane (Fig.~\ref{fig:charmoniumveto}). 
This veto removes charmonium events not only with reconstructed 
$m_{\ell^+\ell^-}$ values 
near the nominal charmonium masses, but also events 
that lie
further away in $m_{\ell^+\ell^-}$ due to photon radiation (more 
pronounced in electron channels) 
or track mismeasurement. 
Removing all of 
these events simplifies the description of the background shape.
Charmonium events can, however, pass this veto if one of the leptons (typically
a muon) and 
the kaon are misidentified as each other.
If reassignment of particle types results in a dilepton mass
consistent with the $J/\psi$ or $\psi(2S)$ mass, the candidate
is vetoed.
There is also significant feed-up from
\mbox{$B\to J/\psi K$} and \mbox{$B\to \psi(2S) K$} into
\mbox{$B \to K^\ast \ell^+ \ell^-$}, 
since energy lost due to bremsstrahlung in $B\to J/\psi K$ 
can be compensated for by including a random
pion. 
If the $K\ell^+\ell^-$ system in a $B\to K^*\ell^+\ell^-$ candidate is
kinematically consistent with $B\to J/\psi(\to\ell^+\ell^-\gamma) K$, assuming
that the photon (which is not directly observed)
was radiated along the direction of either lepton, then the 
candidate is vetoed. Apart from the charmonium vetoes, we analyze the full
$m_{\ell^+\ell^-}$ range.

Continuum background from non-resonant $e^+e^-\to q\overline q$ production
is suppressed using a Fisher 
discriminant~\cite{bib:Fisher}, a linear combination of the input
variables with optimized coefficients. 
The variables are $R_2$; $\cos \theta_B$, the cosine of the angle between 
the $B$ candidate and the beam axis in the c.m.~frame; $\cos \theta_{\rm T}$, 
the cosine of the angle between the thrust axis of the candidate $B$ meson
daughter particles and that of the rest of 
the particles in the c.m.~frame; and $m_{K\ell}$, the invariant mass of the $K$-lepton
system, where the lepton is selected according to
its charge relative to the strangeness 
of the $K^{(*)}$.
The variable $m_{K\ell}$ helps discriminate against background from 
semileptonic $D$ decays, for which $m_{K\ell}<m_D$.  

Combinatorial background from $\BB$ events is suppressed 
using a signal-to-$\BB$ likelihood ratio that  
combines candidate $B$ and dilepton vertex probabilities; 
the significance of the 
dilepton separation along the beam direction; 
$\cos\theta_B$; and the missing energy, $E_{\rm miss}$, of the 
event in the c.m.~frame. The variable $E_{\rm miss}$ provides the
strongest discrimination against $\BB$ background, since
events with semileptonic decays usually have significant unobserved energy
due to neutrinos. For each final state, we select 
at most one combination of particles per event as a $B$ signal
candidate.  If multiple candidates occur, we select the 
candidate with the greatest
number of drift chamber and SVT hits on the charged tracks.

\begin{figure}[!b]
 \begin{center}
   \includegraphics[width=\linewidth]{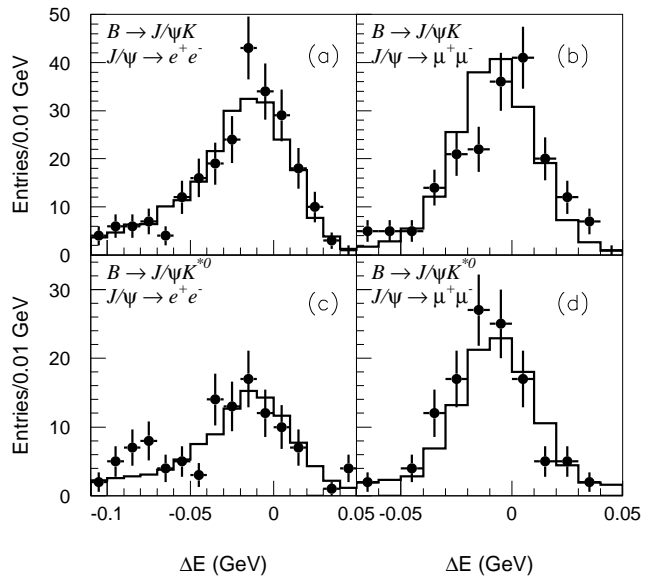}
  \end{center}
  \vspace{-0.5cm}
\caption[Comparison of $\Delta E$ shapes in the charmonium control sample]
{\label{fig:jpside}
 Comparison of event yields and $\Delta E$ shapes between data and Monte Carlo for the charmonium control samples.
The points with error bars show the data, and the solid histograms show the 
prediction of the charmonium Monte Carlo. All of the analysis selection criteria 
have been applied except for the charmonium veto, which is reversed. The large tails
in the $e^+e^-$ modes are due to photon radiation.
Small shifts between data and Monte Carlo are taken into account as
systematic uncertainties on the signal yields.
}
\end{figure}

We use the known charmonium decays 
$B\to J/\psi K^{(*)}$ and $B \to \psi(2S) K^{(*)}$
to check the efficiency of our analysis cuts.
Figure~\ref{fig:jpside} compares
the $\Delta E$ distributions (absolutely normalized) 
of these charmonium samples in 
Monte Carlo with data. 
We find good agreement in both the normalization and the shape.

\begin{table*}[t!]
\caption{
\label{tab:result}
Results from the fits to $B\to K^{(\ast)} \ell^+\ell^-$ and $B\to K^{(*)}e^{\pm}\mu^{\mp}$
modes. 
The columns from left to right are
fitted signal yield~\cite{bib:errors}; 
upper limit on the signal yield; the contribution
of the background to the error on the signal yield, expressed as an effective background yield (see text);
the signal efficiency, $\epsilon$ (not including the branching 
fractions for $K^*$, $K^0$, and $K_S^0$ decays);
the systematic error on the selection efficiency, $(\Delta {\mathcal B}/{\mathcal B})_{\epsilon}$; 
the systematic error from the fit, $(\Delta {\mathcal B}/{\mathcal B})_{\rm fit}$;
the branching 
fraction central value (${\mathcal B}$); and the 
upper limit on the branching fraction, including systematic errors.  
}
\begin{center}
\begin{tabular}{lD{.}{.}{2.5}ccD{.}{.}{2.1}D{.}{.}{3.1}D{.}{.}{3.1}D{.}{.}{2.5}D{.}{.}{2.1}}
 \hline\hline
\multicolumn{1}{c}{Mode}                & \multicolumn{1}{c}{Signal}  & 90\% C.L. &  Effective & \multicolumn{1}{c}{$\epsilon$}  &  \multicolumn{1}{c}{$(\Delta {\mathcal B}/{\mathcal B})_{\epsilon}$}  & \multicolumn{1}{c}{$(\Delta {\mathcal B}/{\mathcal B})_{\rm fit}$}  & \multicolumn{1}{c}{${\mathcal B}/10^{-6}$}  & \multicolumn{1}{c}{${\mathcal B}/10^{-6}$} \\
                                        &   \multicolumn{1}{c}{yield}      & yield     &  background &  \multicolumn{1}{c}{(\%)}       &    \multicolumn{1}{c}{(\%)}            & \multicolumn{1}{c}{(\%)}              &                 &   \multicolumn{1}{c}{90\% C.L.} \\
 \hline 
$B^+\to K^+e^+e^-$                         &  -0.2^{+1.5}_{-0.0}    &  3.1      &   0.7  & 17.5        &   \pm7.6              &  \pm4.0       &  0.0^{+0.4}_{-0.0}  &  0.8  \\    
$B^+\to K^+\mu^+\mu^-$                     &  -0.3^{+1.3}_{-0.0}    &  2.6      &   0.6  & 10.5        &   \pm7.5              &  \pm4.0       & -0.1^{+0.5}_{-0.0}  &  1.2  \\
$B^0\to K^{*0}e^+e^-$                      &   3.8^{+3.8}_{-2.1}    &  8.8      &   1.4  & 10.2        &   \pm8.8              &  \pm11.9      &  2.5^{+2.5}_{-1.4}  &  6.6  \\
$B^0\to K^{*0}\mu^+\mu^-$                  &  -0.3^{+1.7}_{-0.0}    &  3.5      &   0.7  & 8.0         &   \pm10.8             &  \pm3.0       & -0.2^{+1.4}_{-0.0}  &  3.2  \\
$B^0\to K^0 e^+e^-$                        &   1.1^{+2.7}_{-0.9}    &  4.2      &   0.2  & 15.7        &   \pm8.8              &  \pm9.5       &  0.9^{+2.2}_{-0.8}  &  3.9  \\
$B^0\to K^0 \mu^+\mu^-$                    &   0.0^{+1.2}_{-0.0}    &  2.5      &   0.1  & 9.6         &   \pm8.8              &  \pm3.0       &  0.0^{+1.6}_{-0.0}  &  3.7  \\
$B^+\to K^{*+}e^+e^-$                      &  -0.4^{+1.9}_{-0.0}    &  3.8      &   1.6  & 8.5         &   \pm11.0             &  \pm5.0       & -0.8^{+4.3}_{-0.0}  &  9.6  \\
$B^+\to K^{*+}\mu^+\mu^-$                  &   1.2^{+2.4}_{-1.0}    &  4.5      &   0.3  & 5.8         &   \pm13.0             &  \pm7.6       &  3.9^{+8.1}_{-3.2}  & 17.3  \\
\hline
$B^+\to K^+e^{\pm}\mu^{\mp}$               &  -0.4^{+1.4}_{-0.0}    &  2.9      &   1.3  & 16.8        &   \pm5.7              &  \pm4.0       & -0.1^{+0.4}_{-0.0}  &  0.8  \\
$B^0\to K^{*0}e^{\pm}\mu^{\mp}$            &   1.1^{+3.3}_{-1.6}    &  5.3      &   2.7  & 11.9        &   \pm7.1              &  \pm10.4      &  0.6^{+1.8}_{-0.9}  &  3.3  \\
$B^0\to K^0e^{\pm}\mu^{\mp}$               &   1.1^{+2.1}_{-0.9}    &  4.1      &   0.5  & 14.6        &   \pm7.3              &  \pm11.2      &  0.9^{+1.9}_{-0.8}  &  4.1  \\
$B^+\to K^{*+}e^{\pm}\mu^{\mp}$            &  -0.4^{+1.8}_{-0.0}    &  3.5      &   1.1  & 9.3         &   \pm9.6              &  \pm3.0       & -0.8^{+3.8}_{-0.0}  &  8.0  \\
 \hline \hline
\end{tabular}
\end{center}
\end{table*}

\begin{figure}[!tb]
 \begin{center}
   \includegraphics[width=\linewidth]{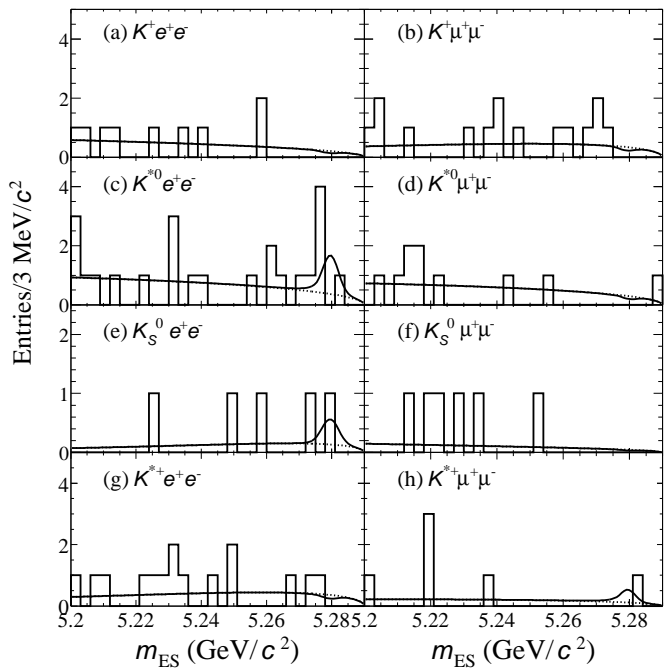}
  \end{center}
  \vspace{-0.3cm}
\caption[Data fit results]
{\label{fig:datafits}
Projections from individual maximum likelihood fits 
onto $m_{\rm ES}$ for the $\Delta E$ signal regions: 
$-0.11 < \Delta E<0.05\ {\rm GeV}$ (electrons)
and $-0.07 < \Delta E<0.05\ {\rm GeV}$ (muons).
The dotted lines show the background component,
and the solid lines show the sum of background and signal components.
}
\end{figure}

We extract the signal and background yields in each channel
using a two-dimensional extended unbinned maximum 
likelihood fit in the region 
defined by \mbox{$m_{\rm ES}$ $>$ 5.2 GeV/$c^2$} and
\mbox{$|\Delta E|$ $<$ 0.25 GeV}. 
The signal shapes,
including the effects of radiation on the $\Delta E$
distribution and the correlation between $m_{\rm ES}$
and $\Delta E$,  
are obtained by parametrizing the
{\tt GEANT}3 Monte Carlo~\cite{bib:GEANT} simulation of the signal.
The background is described by a
function~\cite{bib:bkgshape} with two parameters that are determined
in our fits to the data.
Backgrounds from $\BB$ that peak in the signal region
are suppressed to less than 0.2 events in each mode.
Although we allow the signal yield to be negative,
we have imposed a lower cut-off such that 
the total fit function is positive.
The fit results are shown in Fig.~\ref{fig:datafits}
and summarized in
Table~\ref{tab:result}. We observe no significant signals.

To determine 90\% C.L. upper limits on the signal yields, we generate
and fit a series of toy Monte Carlo samples in which the background
probability density function is taken from our fit to the data,
but the mean number of signal events is varied. We generate ten thousand
samples for each mean value, increasing the mean
until 90\% of the fits to a set of samples 
give a signal yield greater than that obtained by fitting the data.
To give a measure of the sensitivity of the analysis we
list in Table~\ref{tab:result} an effective background yield.
This quantity is
defined as the square of the error
on the signal yield from a fit to a toy Monte Carlo sample
drawn from the background probability function, with no
signal contribution.


Table~\ref{tab:result} lists the systematic uncertainties from the
fit, $(\Delta {\cal B}/{\cal B})_{\rm fit}$, expressed according to their
effect on the limits.
The sensitivity of the limits to the values used for signal-shape
parameters is determined by performing alternative fits 
using parameters from the
$B\to J/\psi K^{(*)}$ control samples. For modes with electrons, 
we also varied the fraction
of signal events in the tail of the $\Delta E$ distribution. 
To determine whether a more general background shape 
would lead to different results, we introduced additional
parameters and allowed for a correlation between
$m_{\rm ES}$ and $\Delta E$. 
This procedure shifted the upper limits by
2\% to 5\%, depending on the mode.
Most of the uncertainty associated with the background shape is 
incorporated in the statistical error on the yield because the background
shape is determined from the fit.

The systematic uncertainties on the efficiency, 
$(\Delta {\cal B}/{\cal B})_{\epsilon},$ are 
listed in Table~\ref{tab:result} and
arise from  
charged-particle tracking ($\pm 1.2\%$/lepton, 
$\pm 2.0\%$ for the pion from $K^*\to K\pi$, and $\pm 1.3\%$/track for
other charged hadrons), 
particle identification ($\pm 1.4\%$/electron,
$\pm 1.0\%$/muon, $\pm 2.0\%$/track for kaons and pions), 
the continuum suppression cut ($\pm 2.0\%$),
the $\BB$ suppression cut ($\pm 3.0\%$), 
$K_S^0$ selection ($\pm 4.0\%$),
Monte Carlo signal statistics ($\pm 3.0\%$ to $\pm 5.0\%$), 
the theoretical model dependence of the efficiency ($\pm 4.0\%$ to $\pm 7.0\%$, depending
on the mode), and
the number of $\BB$ events ($\pm 1.6\%$).
The uncertainties on the efficiencies due to model-dependence of 
form factors are taken to be the full range of variation obtained 
from different theoretical models~\cite{bib:TheoryA}. 
In setting an upper limit, the systematic uncertainties from the efficiency,
$(\Delta {\cal B}/{\cal B})_{\epsilon}$,
and from the fit, $(\Delta {\cal B}/{\cal B})_{\rm fit}$,
are added in quadrature, and the limit is increased by this
factor.


Table~\ref{tab:result} also includes the results
for the lepton-family-violating decays $B\to K^{(*)}e\mu$, where the
signal efficiencies were determined from phase-space Monte Carlo simulations.
We observe no evidence for these decays.

We determine the 
branching fractions 
${\cal B}(B\to K\ell^+\ell^-)$ and ${\cal B}(B\to K^*\ell^+\ell^-)$
averaged over both $B$ meson charge and lepton type ($e^+e^-$ and $\mu^+\mu^-$) 
by performing a simultaneous maximum likelihood fit to the four contributing channels
in each case.
In combining 
the $B\to K^*\ell^+\ell^-$ modes, 
the ratio of branching fractions 
${\cal B}(B\to K^*e^+e^-)/{\cal B}(B\to K^*\mu^+\mu^-)=1.2$ from the model
of Ali {\it et al.}~\cite{bib:TheoryA} is used to
weight the yield in the muon channel relative to that in the electron channel. The
extracted yield corresponds to the electron mode.
The combined fits give
\begin{eqnarray}
{\mathcal B} (B\to K\ell^+\ell^-) & = &(-0.06^{+0.24}_{-0.00}\pm 0.03)\times 10^{-6},   \nonumber\\
{\mathcal B} (B\to K^*\ell^+\ell^-) & = &(0.9^{+1.3}_{-0.9}\pm 0.1)\times 10^{-6}  \nonumber,
\end{eqnarray}
where the first error is statistical and the second is systematic.
We evaluate the upper limits
on these combined modes and obtain 
\begin{eqnarray}
{\mathcal B} (B\to K\ell^+\ell^-) & < &0.50\times 10^{-6}  \  {\rm at}\ 90\%  \ {\rm C.L.}\ \nonumber\\
{\mathcal B} (B\to K^*\ell^+\ell^-) & < &2.9\times 10^{-6} \   {\rm at}\ 90\%  \ {\rm C.L.}\nonumber
\end{eqnarray}

These limits represent an improvement over previously published results from CDF~\cite{bib:CDF}
and CLEO~\cite{bib:CLEO}.
The Belle~\cite{bib:Belle} experiment has also recently 
obtained results on these modes.
We see no evidence for a signal, and our limits are close to many of the predictions based on the 
Standard Model. 
With the rapidly
increasing size of our data sample, we expect to have significantly better sensitivity to these
modes in the future.


We are grateful for the excellent luminosity and machine conditions
provided by our \pep2\ colleagues.
The collaborating institutions wish to thank 
SLAC for its support and kind hospitality. 
This work is supported by
DOE
and NSF (USA),
NSERC (Canada),
IHEP (China),
CEA and
CNRS-IN2P3
(France),
BMBF
(Germany),
INFN (Italy),
NFR (Norway),
MIST (Russia), and
PPARC (United Kingdom). 
Individuals have received support from the Swiss NSF, 
A.~P.~Sloan Foundation, 
Research Corporation,
and Alexander von Humboldt Foundation.



\begin{thebibliography}{99}
\bibitem{bib:bkgshape}{We parametrize the background shape using  \\   
$f(m_{\rm ES},\Delta E)=Ne^{-s\Delta E}m_{\rm ES}\sqrt{1-{m_{\rm ES}^2\over 
    E^{*2}_{\rm b}}}e^{{-\xi\Big(1-{m_{\rm ES}^2\over E^{*2}_{\rm b}}\Big)}},$
where $N$ is a normalization factor and $s$ and $\xi$ are free parameters
determined from the fit to the data. 
}
\bibitem{bib:errors}{
Whenever possible, we report two-sided 68\% central confidence intervals.
For channels constrained by the
requirement that the total fit function be non-negative, we quote a
single-sided 68\% confidence interval and set the lower statistical
error to zero.
}
\end{thebibliography}
\end{document}